# Derivation of the Time-Dependent Schrödinger Equation from the Time-Independent One

Nikolay A. Vinokurov[1,2,*]

[1]Budker Institute of Nuclear Physics, Siberian Brunch of Russian Academy of Sciences, pr. Akad. Lavrent'eva 11, Novosibirsk, 630090, Russia

[2]Novosibirsk State University, ul. Pirogova 1, Novosibirsk 630090, Russia

ABSTRACT

A derivation of the time-dependent Schrödinger equation from the time-independent one is considered. Instead of time, the coordinate of an additional degree of freedom, the clock, is introduced into the original time-independent Schrödinger equation. It is shown that the standard time-dependent Schrödinger equation can be obtained for the semiclassical clock only. For elucidation of the physical meaning of the equation obtained in this way, various types of clocks are discussed. In addition, the corresponding equation for the density matrix and formulas for the mean values of operators are derived.

*Introduction.* In his first article "Quantization as an eigenvalue problem" [1], Schrödinger derived an equation for finding energy levels, the time-independent Schrödinger equation (SE), which made it possible to describe the emission spectrum of a hydrogen atom. After that, with the help of various improvements of the Hamiltonian of the time-independent SE, a detailed description of the emission and absorption spectra under different conditions was obtained (for example, the Zeeman and Stark effects, the fine and hyperfine structure of lines, etc.). It is worth noting that the problem of measurements with the help of macroscopic instruments practically does not concern the interpretation of spectra. Therefore, the well-known remarks [2] about the incompleteness of quantum mechanics do not apply to the results obtained with the use of the time-independent SE. In addition, spectra have nothing to do with interpretations of quantum mechanics and, in particular, with the concept of probability. Mathematically, this is expressed in the fact that the spectrum of the energy operator can be found without mentioning of the wave function, as was done by Pauli and Dirac [3, 4]. In this sense, we can say that the correctness of the stationary SE is directly confirmed by the experimental results and this part of quantum mechanics is complete.

Transition from the time-independent SE to the time-dependent one is nontrivial (see, for example, [5, 6]). This is partly due to the introduction of time, which is different from coordinates in many ways, into the equation.

In classical mechanics, time is often replaced by the coordinate of an additional degree of freedom (clock), and the law of motion is replaced by a description of the trajectory in an extended coordinate space that includes the coordinate of the clock. In this case, a dynamic problem is reduced to a geometric one. This corresponds to the usual perception of time, associated with the motion of the Earth. Therefore, it seems natural to consider the same approach in quantum mechanics, that is, to introduce a clock into the time-independent SE.

*The simplest clock.* Consider the time-independent SE for an isolated conservative system consisting of some subsystem with the generalized coordinates $\mathbf{q} = (q_1, q_2, \ldots q_N)$ and the Hamiltonian $H(p_1, p_2, \ldots p_N, q_1, q_2, \ldots q_N, q_c)$ and a clock with the Hamiltonian $H_c(p_c, q_c)$:

$$(H + H_c)\Psi = E\Psi, \qquad (1)$$

where $\Psi$ is the wave function of the system, $E$ is the total energy of the system, and the simplest clock is a non-relativistic particle with the coordinate ("clock hand") $q_c$ and mass $M$, i.e.,

$$H_c = -\frac{\hbar^2}{2M}\frac{\partial^2}{\partial q_c^2}. \qquad (2)$$

Then

$$\frac{\hbar^2}{2M}\frac{\partial^2}{\partial q_c^2}\Psi = (H - E)\Psi. \qquad (3)$$

If $H$ does not depend on $q_c$, the separation of variables $\Psi = \psi_c(q_c)\psi(\mathbf{q})$ gives

$$\frac{\hbar^2}{2M\psi_c}\frac{\partial^2}{\partial q_c^2}\psi_c = \frac{(H-E)\psi}{\psi} = -E_c. \qquad (4)$$

$$H\psi_n = (E - E_c)\psi_n = E_n\psi_n$$

Then

$$\psi_c \propto \exp\left(\pm i\frac{q_c}{\hbar}\sqrt{2ME_c}\right), \qquad (5)$$

$$E_c = E - E_n, \tag{6}$$

and the complete integral (solution) of Eq. (3) is

$$\Psi = \sum_n \psi_n(\mathbf{q}) \left\{ A_n \exp\left[i \frac{q_c}{\hbar} \sqrt{2M(E-E_n)}\right] + B_n \exp\left[-i \frac{q_c}{\hbar} \sqrt{2M(E-E_n)}\right] \right\}, \tag{7}$$

where the sum over $n$ for the continuous part of the spectrum of $H$ means the corresponding integration.

The respective solution for the "initial-value problem" is

$$\Psi(\mathbf{q}, q_c) =$$
$$\sum_n \int \Psi(\mathbf{q}',0) \psi_n^*\, d\mathbf{q}' \, \psi_n(\mathbf{q}) \cos\left[\frac{q_c}{\hbar} \sqrt{2M(E-E_n)}\right] + \tag{8}$$
$$\sum_n \int \frac{\partial \Psi}{\partial q_c}(\mathbf{q}',0) \psi_n^*\, d\mathbf{q}' \, \psi_n(\mathbf{q}) \frac{\hbar}{\sqrt{2M(E-E_n)}} \sin\left[\frac{q_c}{\hbar} \sqrt{2M(E-E_n)}\right]$$

where the normalization $\int \psi_m(\mathbf{q}) \psi_n^*(\mathbf{q}) d\mathbf{q} = \delta_{mn}$ is used. According to Eq. (8), the value $\operatorname{Im} \int \Psi^* \partial \Psi/\partial q_c\, d\mathbf{q}$ does not depend on $q_c$.

For special initial conditions $B_n = 0$ (which means that the clock is moving in the positive direction of the $q_c$ axis), or

$$\int \frac{\partial \Psi}{\partial q_c}(\mathbf{q}',0) \psi_n^*\, d\mathbf{q}' = i \frac{\sqrt{2M(E-E_n)}}{\hbar} \int \Psi(\mathbf{q}',0) \psi_n^*\, d\mathbf{q}', \tag{9}$$

$$\Psi(\mathbf{q}, q_c) = \sum_n \int \Psi(\mathbf{q}',0) \psi_n^*\, d\mathbf{q}' \, \psi_n(\mathbf{q}) \exp\left[i \frac{q_c}{\hbar} \sqrt{2M(E-E_n)}\right]. \tag{10}$$

If the coefficients $A_n$ in Eq. (7) are non-zero only for $E_n \ll E$, then

$$i\hbar \frac{\partial}{\partial q_c}\left( \Psi \exp \frac{-i q_c \sqrt{2ME}}{\hbar} \right) =$$
$$\exp \frac{-i q_c \sqrt{2ME}}{\hbar} \sum_n \int \Psi(\mathbf{q}',0) \psi_n^*\, d\mathbf{q}' \, \psi_n(\mathbf{q}) \left[\sqrt{2ME} - \sqrt{2M(E-E_n)}\right] \exp \frac{i q_c \sqrt{2M(E-E_n)}}{\hbar} \approx . \tag{11}$$
$$\sqrt{\frac{M}{2E}} H\left(\Psi \exp \frac{-i q_c \sqrt{2ME}}{\hbar}\right)$$

Equation (11) looks like the time-dependent SE with the "time"

$$t = q_c \sqrt{\frac{M}{2E}}, \tag{12}$$

corresponding to the velocity of the clock $\sqrt{2E/M}$.

If $H$ depends on $q_c$, one can use a more standard way to derive Eq. (11). The substitution

$$\Psi = \psi(\mathbf{q}, q_c) \exp \frac{i\sqrt{2ME} q_c}{\hbar} \tag{13}$$

gives

$$i\hbar \sqrt{\frac{2E}{M}} \frac{\partial}{\partial q_c} \psi + \frac{\hbar^2}{2M} \frac{\partial^2}{\partial q_c^2} \psi = H\psi, \tag{14}$$

or

$$i\hbar \sqrt{\frac{2E}{M}} \frac{\partial}{\partial q_c} \left( \psi - \frac{i\hbar}{2\sqrt{2ME}} \frac{\partial}{\partial q_c} \psi \right) = H\psi. \tag{15}$$

At

$$\left| \frac{\partial}{\partial q_c} \ln \psi \right| \ll \frac{2\sqrt{2ME}}{\hbar}, \tag{16}$$

that is, with a sufficiently large clock momentum, we can neglect the second term in the brackets at the left side of Eq. (15) and obtain the "time-dependent" SE

$$i\hbar \sqrt{\frac{2E}{M}} \frac{\partial}{\partial q_c} \psi = H\psi \tag{17}$$

with the "time" given by Eq. (12). The condition of Eq. (16) means that the localization of the clock $\Delta q_c$ significantly exceeds its de Broglie wavelength, that is, the clock is semiclassical ("macroscopic"). In this case, the total wave function $\Psi$ given by Eq. (13) is the product of a slow-varying complex amplitude $\psi$ and a fast-oscillating function of the clock coordinate. If a rigid rotator is used as a clock, then the coordinate $q_c$ is the angle of rotation, the moment of inertia should be substituted for the mass, and the speed of the clock $\sqrt{2E/M}$ is the angular frequency of rotation.

*Longitudinal motion as a clock.* An illustrative example of replacing time with the "clock" coordinate is the paraxial motion of a high-energy nonrelativistic particle along the $z$ axis [2]. The Hamiltonian has the form

$$H_{tot} = \frac{p_x^2 + p_y^2}{2m} + U(x,y,z) + \frac{p_z^2}{2m} = H(p_x, p_y, x, y, z) + \frac{p_z^2}{2m}. \tag{18}$$

In this case, instead of Eq. (3) we have

$$\frac{\hbar^2}{2m} \frac{\partial^2 \Psi}{\partial z^2} = (H - E)\Psi, \tag{19}$$

and replacement of Eq. (13) with a sufficiently large longitudinal momentum $\sqrt{2mE}$ gives instead of Eq. (17)

$$i\hbar \sqrt{\frac{2E}{m}} \frac{\partial}{\partial z} \psi = H\psi, \tag{20}$$

which differs from Eq. (45.8) of [2] only in the fact that $H$ includes not only the potential energy, but also the kinetic energy of the transverse degrees of freedom.

Such an implicit replacement of the time by the semiclassical value of the longitudinal coordinate is always done when electron diffraction on slits (including modifications, for example, the Aharonov–Bohm effect) or the Stern–Gerlach experiment is described [2, 5–7]. This is natural, since the screens and deflecting magnets do not move, and the projection of the particle momentum onto the longitudinal coordinate is almost constant.

In the theory of diffraction, an equation of the type of Eq. (20) is called a parabolic equation [8]. It is derived from the three-dimensional Helmholtz equation for the paraxial propagation of short wavelength radiation. One way to derive it is to replace the Green function of the Helmholtz equation (the field of a point source) $\exp\left(ik\sqrt{x^2+y^2+z^2}\right)/\sqrt{x^2+y^2+z^2}$ with its approximation $\exp\left[ikz + ik(x^2+y^2)/(2z)\right]/z$ at $\sqrt{x^2+y^2} \ll z$, that is, to use the Fresnel diffraction approximation.

Note that transforming time-independent SE (1), we considered a time-independent scattering problem with specific boundary conditions given on a remote closed surface. In this case, the motion of the system was considered to be paraxial with respect to the coordinate chosen as the clock, and the "direction of time" is given by the choice of the sign of the momentum of the high-energy degree of freedom used as the clock.

*A more general one-dimensional semiclassical system.* The above derivation is easy to generalize. Considering in Eq. (1) the Hamiltonian $H$ of the initial system as a small perturbation, we can write down the semiclassical solution of the unperturbed equation in the form $\left[p_0(q_c)\right]^{-1/2} \exp\left[i\int_0^{q_c} p_0(q)dq/\hbar\right]$, where $H_c(p_0, q_c) = E$, and with the substitution

$$\Psi = \frac{\psi(\mathbf{q}, q_c)}{\sqrt{p_0(q_c)}} \exp \frac{i\int_0^{q_c} p_0(q)dq}{\hbar}, \tag{21}$$

we get for

$$H_c = \sum_{n=0}^{\infty} h_n(q_c)(-i\hbar)^{2n} \frac{\partial^{2n}}{\partial q_c^{2n}} \tag{22}$$

$H_c \Psi \approx$

$$\left\{ E - \frac{i\hbar}{2} \frac{\partial p_0(q_c)}{\partial q_c} p_0 \frac{\partial}{\partial p_c} \left[ \frac{1}{p_c} \frac{\partial H_c(p_c, q_c)}{\partial p_c} \right] \right\} \frac{\psi}{\sqrt{p_0(q_c)}} \exp \frac{i\int_0^{q_c} p_0(q)dq}{\hbar} - \tag{23}$$

$$\frac{i\hbar}{\sqrt{p_0(q_c)}} \exp \frac{i\int_0^{q_c} p_0(q)dq}{\hbar} \frac{\partial H(p_c, q_c)}{\partial p_c} \frac{\partial \psi}{\partial q_c}$$

Then Eq. (1) gives

$$H\psi = \sqrt{p_0(q_c)} \exp\frac{-i\int_0^{q_c} p_0(q)dq}{\hbar}(E - H_c)\exp\frac{i\int_0^{q_c} p_0(q)dq}{\hbar}\frac{\psi}{\sqrt{p_0(q_c)}} \approx$$
$$i\hbar\frac{\partial H_c(p_c,q_c)}{\partial p_c}\bigg|_{p_c=p_0(q_c)}\frac{\partial \psi}{\partial q_c} + \frac{i\hbar}{2}\frac{\partial p_0(q_c)}{\partial q_c}p_0\frac{\partial}{\partial p_c}\left[\frac{1}{p_c}\frac{\partial H_c(p_c,q_c)}{\partial p_c}\right]\psi \quad (24)$$

If the clock speed changes rather slowly, then, neglecting the second term on the right-hand side of (24), we obtain

$$H\psi \approx i\hbar\frac{\partial H_c(p_c,q_c)}{\partial p_c}\bigg|_{p_c=p_0(q_c)}\frac{\partial \psi}{\partial q_c}. \quad (25)$$

Note that for the momentum-quadratic clock Hamiltonian $H_c = p_c^2/(2M) + U(q_c)$, the second term on the right-hand side of Eq. (24) is equal to zero and

$$H\psi \approx i\hbar\sqrt{2\frac{E-U(q_c)}{M}}\frac{\partial \psi}{\partial q_c} = i\hbar\frac{\partial \psi}{\partial \tau}, \quad (26)$$

that is, the clock is uneven, but for convenience, one can introduce a special variable

$$\tau = \sqrt{\frac{M}{2}}\int\frac{dq_c}{\sqrt{E-U(q_c)}}. \quad (27)$$

*Harmonic oscillator.* For a clock in the form of a harmonic oscillator, one can write its Hamiltonian $H_c = p_c^2/(2M) + M\omega^2 q_c^2/2$ in the coherent state representation (see, for example, [9]). Introducing the raising

$$a^+ = \frac{q_c\sqrt{M\omega} - ip_c/\sqrt{M\omega}}{\sqrt{2\hbar}} \quad (28)$$

and lowering

$$a = \frac{q_c\sqrt{M\omega} + ip_c/\sqrt{M\omega}}{\sqrt{2\hbar}} \quad (29)$$

operators, we get $\left[a, a^+\right] = 1$ and

$$H_c = \frac{\hbar\omega}{2}(a^+a + aa^+) = \hbar\omega\left(a^+a + \frac{1}{2}\right). \quad (30)$$

Let $\alpha$ be an eigenvalue of the lowering operator $a$. Then in the coherent state representation ($\alpha$-representation)

$$a^+ = -\frac{\partial}{\partial \alpha}, \quad (31)$$

and

$$H_c = -\hbar\omega\left(\alpha\frac{\partial}{\partial \alpha} + \frac{1}{2}\right). \quad (32)$$

Taking into account Eq. (32), we obtain instead of Eq. (3)

$$\hbar\omega\frac{\partial}{\partial \ln\alpha}\Psi = \left(H - E - \frac{\hbar\omega}{2}\right)\Psi, \quad (33)$$

and substitution

$$\Psi = \psi(\mathbf{q}, q_c)\alpha^{-E/(\hbar\omega)-1/2} \quad (34)$$

gives

$$\hbar\omega\frac{\partial}{\partial \ln\alpha}\psi = H\psi. \quad (35)$$

When deriving Eq. (35), we assumed that $H$ does not depend on $q_c$. Taking into account the fact that in the Heisenberg representation the rate of change of the operator $a$ is equal to

$$\dot{a} = \frac{i}{\hbar}[H_c, a] = -i\omega a, \quad (36)$$

one can define the variable $\tau = i\ln\alpha/\omega$. Then Eq. (35) takes the standard form

$$i\hbar\frac{\partial \psi}{\partial \tau} = H\psi. \quad (37)$$

Equation (37) is exact, but the variable $\tau$ describing the state of the clock is complex.

*Mixed states.* When projection operators or, more generally, density matrices $R$, are used, the time-independent SE leads to

$$R(H+H_c) = (H+H_c)R = ER, \tag{38}$$

or

$$[H,R] = -[H_c,R] \tag{39}$$

and

$$H_c R = (E-H)R. \tag{40}$$

The right-hand side of Eq. (39) can be written in the coordinate representation [2] as

$$[H,R] = -\left[H_c\left(-i\hbar\frac{\partial}{\partial q_c}, q_c\right) - H_c^*\left(-i\hbar\frac{\partial}{\partial q_c'}, q_c'\right)\right]R(\mathbf{q},q_c,\mathbf{q}',q_c'). \tag{41}$$

Taking for simplicity the clock Hamiltonian (Eq. (2)) and making the substitution

$$R(\mathbf{q},q_c,\mathbf{q}',q_c') = \exp\frac{i\sqrt{2ME}(q_c - q_c')}{\hbar} R_s(\mathbf{q},q_c,\mathbf{q}',q_c'), \tag{42}$$

similar to Eq. (13), which singles out a rapidly oscillating factor, at high energy of the clock we obtain

$$[H, R_s(\mathbf{q},q_c,\mathbf{q}',q_c')] =$$

$$\exp\frac{i\sqrt{2ME}(q_c' - q_c)}{\hbar}\frac{\hbar^2}{2M}\left(\frac{\partial^2}{\partial q_c^2} - \frac{\partial^2}{\partial q_c'^2}\right)\left\{\exp\frac{i\sqrt{2ME}(q_c - q_c')}{\hbar}R_s(\mathbf{q},q_c,\mathbf{q}',q_c')\right\} \approx . \tag{43}$$

$$i\hbar\sqrt{\frac{2E}{M}}\left\{\frac{\partial}{\partial q_c}R_s(\mathbf{q},q_c,\mathbf{q}',q_c') + \frac{\partial}{\partial q_c'}R_s(\mathbf{q},q_c,\mathbf{q}',q_c')\right\}$$

For $q_c = q'_c$, that is, for the elements of the density matrix $R$ diagonal in $q_c$, from Eq. (43) we obtain a generalization of time-dependent SE (17) to the case of mixed states:

$$i\hbar\sqrt{\frac{2E}{M}}\frac{\partial}{\partial q_c}\rho(\mathbf{q},\mathbf{q}',q_c) = [H,\rho(\mathbf{q},\mathbf{q}',q_c)], \tag{44}$$

where $\rho(\mathbf{q},\mathbf{q}',q_c) = R(\mathbf{q},q_c,\mathbf{q}',q_c)$. It is worth noting that Eq. (43) is more general than Eq. (44), as it describes two-time correlations.

On the other hand,

$$R_m(\mathbf{q},q_c,\mathbf{q}',q_c') =$$
$$\delta(q_{c0} - q_c)\delta(q_{c0} - q_c')\int\delta(q_{c0} - x)\delta(q_{c0} - x')R(\mathbf{q},x,\mathbf{q}',x')dxdx' = \tag{45}$$
$$\delta(q_{c0} - q_c)\delta(q_{c0} - q_c')\rho(\mathbf{q},\mathbf{q}',q_{c0})$$

is the density matrix of the system after measurement of the clock coordinate $q_c$ with the result $q_{c0}$ (see, for example, [10]). Then one can say that $\rho(\mathbf{q},\mathbf{q}',q_c)$ is the conditional density matrix of the subsystem with the coordinates $\mathbf{q}$ at the clock readings $q_c$.

*Interpretation of the resulting equations.* According to the standard interpretation of the wave function and the density matrix, the mean value of the quantity represented by the function $\varphi(\mathbf{q},q_c,\mathbf{q}',q_c')$ is given by Eq. (2.1) from [2]:

$$\langle\varphi\rangle = \int\varphi(\mathbf{q},q_c,\mathbf{q}',q_c')R(\mathbf{q},q_c,\mathbf{q}',q_c')d\mathbf{q}dq_c d\mathbf{q}'dq_c'. \tag{46}$$

For example, at $\varphi = \delta(\mathbf{q}-\mathbf{q}_0)\delta(q_c - q_{c0})\delta(\mathbf{q}'-\mathbf{q}_0)\delta(q_c' - q_{c0})$, Eq. (46) gives the probability of finding the system at the point $\mathbf{q}_0$ with the clock readings $q_{c0}$:

$$\langle\varphi\rangle = R(\mathbf{q}_0,q_{c0},\mathbf{q}_0,q_{c0}) = \rho(\mathbf{q}_0,\mathbf{q}_0,q_{c0}). \tag{47}$$

To obtain the average value of the operator $f(\mathbf{p},\mathbf{q})$ at the clock readings $q_{c0}$, one should use

$$\varphi = \delta(\mathbf{q}-\mathbf{q}')\delta(q_c - q_{c0})\delta(q_c' - q_{c0})f\left(-i\hbar\frac{\partial}{\partial\mathbf{q}},\mathbf{q}\right). \tag{48}$$

Then

$$\langle\varphi\rangle = \int\delta(\mathbf{q}-\mathbf{q}')f\left(-i\hbar\frac{\partial}{\partial\mathbf{q}},\mathbf{q}\right)R(\mathbf{q},q_{c0},\mathbf{q}',q_{c0})d\mathbf{q}d\mathbf{q}' = \mathrm{Sp}(f\rho). \tag{49}$$

*Conclusion.* Thus, time is introduced into the time-independent SE through the readings of a semiclassical clock. This may indicate the semiclassical nature of the concept of time and, in particular, the semiclassical nature of the time-dependent SE. From the point of view of standard quantum mechanics, the last statement is quite natural, since time is not an observable in quantum mechanics.

We can say that the wave function $\Psi(\mathbf{q}, q_c)$ describes the "relative state" [11] of the subsystem with the coordinates **q** at the measured value of the clock readings $q_c$. Due to the semiclassical nature of the clock, the measurements give a result with a small uncertainty and there is no noticeable "splitting of the worlds".

Technically, the derivation of the time-dependent SE practically repeats, as already noted, the derivation of the parabolic equation of the diffraction theory from the Helmholtz equation. Namely, Eq. (17) describes the paraxial wave along the $q_c$ axis in the extended configuration space ($q_1, q_2, \ldots q_N, q_c$).

*vinokurov@inp.nsk.su

[1] E. Schrödinger, Ann. Physik **79**, 361 (1926).
[2] L. D. Landau and E. M. Lifshitz, *Quantum mechanics* (Pergamon Press, 1991).
[3] W. Pauli, Z. Phys. **36**, 336 (1926).
[4] P. A. M. Dirac, Proc. Roy. Soc. A **109**, 642 (1926).
[5] D. Bohm, *Quantum theory* (Prentice-Hall Inc., 1952).
[6] A. Peres, *Quantum theory: Concepts and Methods* (Kluwer Academic Publishers, 2002).
[7] R. P. Feynman, R. B. Leighton, M. Sands, *The Feynman Lectures on Physics* (Addison-Wesley, 1965).
[8] V. A. Fock, *Electromagnetic Diffraction and Propagation Problems* (Pergamon Press, 1965).
[9] R. Glauber, in *Quantum optics and electronics* Eds C DeWitt, A Blandin, C Cohen- Tannoudji, (Gordon and Breach, Science Publishers, 1965).
[10] A. Sudbery, *Quantum mechanics and the particles of nature* (Cambridge University Press, 1986).
[11] H. Everett III, Rev. Mod. Phys. **29** 3 454 (1957).